\documentclass[aps,pra,superscriptaddress,reprint,showkeys,showpacs]{revtex4-1}
\usepackage{amsmath}    
\usepackage{graphicx}   
\usepackage{array}
\usepackage[dvipdfx,cmyk]{xcolor}     
\usepackage[caption=false]{subfig}  
\usepackage[colorlinks,linkcolor=red,citecolor=brown,urlcolor=blue]{hyperref}
\usepackage{dsfont,bm}

\newcommand{\be}{\begin{equation}}
\newcommand{\ee}{\end{equation}}
\newcommand{\ben}{\begin{eqnarray}}
\newcommand{\een}{\end{eqnarray}}
\newcommand{\bes}{\begin{subequations}}
\newcommand{\ees}{\end{subequations}}
\newcommand{\bF}{\begin{figure}}
\newcommand{\eF}{\end{figure}}

\newcommand{\bra}[1]{\langle{#1}\vert}
\newcommand{\ket}[1]{\vert{#1}\rangle}

\begin{document}

\title{Generation of entanglement in systems of intercoupled qubits}

\author{Levon Chakhmakhchyan}
\affiliation{Laboratoire Interdisciplinaire Carnot de Bourgogne, UMR CNRS 6303 Universit\'{e} de Bourgogne, 21078 Dijon Cedex, France}
\affiliation{Institute for Physical Research, 0203 Ashtarak-2, Armenia}
\affiliation{A.I. Alikhanyan National Science Laboratory, Alikhanian Br. 2, 0036 Yerevan, Armenia}

\author{Claude Leroy}
\affiliation{Laboratoire Interdisciplinaire Carnot de Bourgogne, UMR CNRS 6303 Universit\'{e} de Bourgogne, 21078 Dijon Cedex, France}

\author{Nerses Ananikian}
\affiliation{A.I. Alikhanyan National Science Laboratory, Alikhanian Br. 2, 0036 Yerevan, Armenia}

\author{St\'{e}phane Gu\'{e}rin}
\affiliation{Laboratoire Interdisciplinaire Carnot de Bourgogne, UMR CNRS 6303 Universit\'{e} de Bourgogne, 21078 Dijon Cedex, France}
\date{\today}

\begin{abstract}
We consider systems of two and three qubits, mutually coupled by Heisenberg-type exchange interaction and interacting with external laser fields. We show that these systems allow one to create maximally entangled Bell states, as well as three-qubit Greenberger-Horne-Zeilinger and W states. In particular, we point out that some of the target states are the eigenstates of the initial bare system. Due to this, one can create entangled states by means of pulse area and adiabatic techniques, when starting from a separable (non-entangled) ground state. On the other hand, for target states, not present initially in the eigensystem of the model, we apply the robust stimulated Raman adiabatic passage and $\pi$ pulse techniques, that create desired coherent superpositions of non-entangled eigenstates.
\end{abstract}

\maketitle

\section{Introduction}

Quantum entanglement is one of the vital resources in most applications of quantum information science. It is essential to the implementation of various quantum protocols, including quantum teleportation \cite{teleport, teleport1}, quantum cryptography \cite{QKD, QKD1}, dense coding \cite{dense, dense1}, and is at the heart of quantum computation \cite{comp}.

Different methods for creating entangled states have been proposed recently. In particular, techniques for generating polarization-entangled photon pairs by means of radiative decay of biexcitons of quantum dots \cite{biex, demand}, parametric down conversion \cite{down, prl, federov}, or four-wave mixing processes \cite{four} are known. Furthermore, protocols involving entanglement generation in continuous-variable systems \cite{cont}, and, particularly, making use of quantum memories \cite{memory, me1}, are also widely used for implementing various quantum communication schemes \cite{contprot, contprot1, contprot2}. In addition, schemes for creating atom-photon entangled states \cite{fed}, aimed, e.g., at construction of long-range quantum networks \cite{lr1, lr2}, are under active investigation as well. On the other hand, solid-state systems are considered as natural entanglement resources on their own turn. Namely, the exchange type interaction that couples quantum spins, nested at the sites of a solid's lattice, may give rise to entangled ground and thermal states \cite{entangle, me2, me22, ananik13} (the existence of the latter states have been proven experimentally by means of heat capacity and magnetic susceptibility measurements \cite{exp, exp1}). Additionally, recent experimental observations show a possibility of entangling macroscopic millimeter-sized diamonds at room temperature \cite{diam}.

In this paper we propose another method for generating maximally entangled two-qubit states, and three-qubit Greenberger-Horne-Zeilinger (GHZ) and W states, which are an essential building block for quantum communication and quantum information processing \cite{W}. As is known, in the two-qubit case, all maximally entangled states, known as Bell states, are equivalent (up to local changes of basis). Meanwhile, three-qubit entangled states can be created in two fundamentally different ways, resulting in GHZ-type and W-type states, that cannot be transformed into each other by local operations and classical communication \cite{charac}. Within our approach, the above entangled states are prepared in systems of intercoupled qubits, interacting with incident laser fields. The mutual interqubit interaction is chosen here to be of a Heisenberg-type exchange character. The latter arises in many systems, e.g., coupled semiconductor quantum dots \cite{inter} (as well as in the biexciton system of a single semiconductor quantum dot, that acts as a two-qubit register \cite{inter1, inter2}), superconducting phase and charge qubits \cite{jos, jos1, parano1}, atoms (ions) trapped in a cavity (ion trap) within the dispersive limit \cite{ionent, me3, dispersive3}, etc. A few methods for creating entangled states, using, in particular, qubit rotation and quantum logical operations in similar three-qubit systems \cite{jos2}, as well as rapid adiabatic passage (RAP) with chirped gaussian pulses in two-qubit systems \cite{rapexp} have been reported. Furthermore, protocols for implementing high-speed and high-fidelity single-qubit and C-NOT gates  via microwave fields in coupled superconducting qubits have been proposed \cite{parano}.  On our part, we demonstrate schemes for generating all four Bell states, GHZ and W states by means of adiabatic [stimulated Raman adiabatic passage (STIRAP), fractional STIRAP, RAP] and pulse area techniques, each of which have their own advantages \cite{revrap}. We note that the STIRAP method for generating a specific type of two-qubit entangled states has been described in Refs.~\cite{unanyan, diamond}. However the interqubit coupling there was taken of a separable (diagonal) character, which changes the eigenstate structure of the bare qubit system drastically. On the other hand our method allows one to manipulate the amount of entanglement in a continuous way, fixing, e.g., the area of the incident laser pulse (for quantifying entanglement we use the logarithmic negativity, a measure of entanglement for a bipartite system \cite{negat, negat1}).

The paper is organized as follows: in Sec.~\ref{model} we introduce the model of intercoupled qubits interacting with incident laser fields, and derive its main properties for the case of two and three qubits. In Sec.~\ref{ent} we present schemes for generating Bell states and three-qubit GHZ and W states by means of pulse area and adiabatic techniques. We draw our conclusions in Sec.~\ref{conc}.

\section{Model}\label{model}
We consider a collection of qubits, coupled to one another by means of exchange-type interaction, and also coupled to external laser fields, leading to the Hamiltonian (in units such that $\hbar=1$):
\begin{eqnarray}\label{1}
\begin{split}
&\mathcal{H}=\mathcal{H}_{qq}+\mathcal{H}_{ql}, \\
&\mathcal{H}_{qq}=\lambda\sum_{i\neq j}S_i^+  S_j^- + \frac{1}{2}\sum_{i=1}^N \omega^i_0 S_i^z, \\
&\mathcal{H}_{ql}=-\sum_{i=1}^N  d^i\sum_{j=1}^n E_j(t).
\end{split}
\end{eqnarray}
Here, $\mathcal{H}_{qq}$ is the Hamiltonian of the intercoupled qubits, with $S_i^+=|1_i\rangle\langle0_i|$, $S_i^-=|0_i\rangle\langle1_i|$, $S_i^z=|1_i\rangle\langle1_i|-|0_i\rangle\langle0_i|$ ($|0_i\rangle$ and $|1_i\rangle$ are the ground and excited states respectively, of the $i^\mathrm{th}$ qubit), $\lambda$ is the strength of interqubit coupling and $\omega_0^i$ is the level splitting of the $i^\mathrm{th}$ qubit. The Hamiltonian $\mathcal{H}_{ql}$ represents the interaction of $N$ qubits and $n$ laser fields of the electric field $E_j(t) =
\varepsilon_j(t)e^{-i\omega^j_lt} + \varepsilon_j^*(t)e^{i\omega^j_lt}$, where $\varepsilon_j(t)$ and $\omega^j_l$ are, respectively, the slowly varying envelope and the frequency of the laser ($j=1, 2, ..., n$). Finally, the dipole moment of the $i^\mathrm{th}$ qubit is defined as an operator of the form $d^i=d_{10}^i|1_i\rangle \langle0_i|+d_{01}^i|0_i\rangle \langle1_i|$ with the corresponding matrix elements $d_{kl}^i=\bra{k}d^i\ket{l}$. Hereafter we additionally assume the qubits to have equal level splitting, i.e., $\omega^i_0\equiv\omega_0$ for $i=1, 2, ..., N$.
\subsection{Two intercoupled qubits}
We start with the case of two qubits, coupled to four incident laser fields ($N=2$ and $n=4$). The eigenvectors of $\mathcal{H}_{qq}$ are the following well-known states:
\begin{eqnarray}\label{2}
\begin{split}
&\ket{\psi_{00}}=\ket{00}, \\
&\ket{\psi_{-}}=\frac{1}{\sqrt{2}}(\ket{10}-\ket{01}), \\
&\ket{\psi_{+}}=\frac{1}{\sqrt{2}}(\ket{10}+\ket{01}), \\
&\ket{\psi_{11}}=\ket{11}, \\
\end{split}
\end{eqnarray}
with corresponding eigenenergies: $E_{00}=-\omega_0$, $E_{\pm}=\pm \lambda$ and $E_{11}=\omega_0$. As already mentioned in the Introduction, a similar system was studied in Ref.~\cite{unanyan}, where the interqubit coupling, however, was taken of a diagonal character, i.e., involving only $S_1^zS_2^z+S_2^zS_1^z$ terms. Within this type of interaction the bare system possesses only separable states $\{\ket{00}, \ket{10}, \ket{01}, \ket{11}\}$.

\begin{figure}[h!]
\begin{center}
\includegraphics[width=7cm]{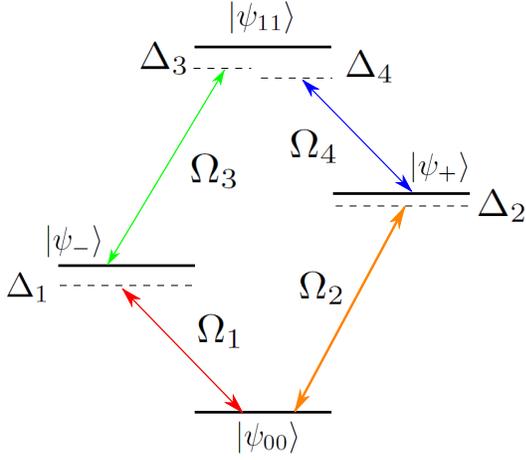}
\caption {(Color online) The effective level scheme for two intercoupled qubits, interacting with four incident laser fields. \label{levels}}
\end{center}
\end{figure}

Expanding the total wavefunction $\ket{\Psi}$ of the system in the basis, given by Eq.~(\ref{2}), and substituting the corresponding expression into the time-dependant Schr\"{o}dinger equation $i d \ket{\Psi}/d t=\mathcal{H} \ket{\Psi}$, we obtain a set of equations for the amplitudes $\{a_{00}(t), a_-(t), a_+(t), a_{11}(t)\}$, which, within the rotating-wave approximation (RWA), reads (see Fig.~\ref{levels} for the definition of detunings):
\begin{eqnarray} \label{3}
&& i\frac{d}{d t} \left( \begin{array}{c} a_{00}(t) \\ a_-(t) \\ a_+(t) \\ a_{11}(t) \\ \end{array} \right)=\\ && \left(
                         \begin{array}{llll}
                           -\Delta_2    & -\Omega_1^*(t)        & -\Omega_2^*(t) & 0 \\
                           -\Omega_1(t) & \Delta_1-\Delta_2    & 0             & -\Omega_3^*(t) \\
                           -\Omega_2(t) & 0                    & 0             & -\Omega_4^*(t) \\
                           0            & -\Omega_3(t)         & -\Omega_4(t)  & \Delta_4 \\
                         \end{array}
                       \right) \cdot \left( \begin{array}{c} a_{00}(t) \\ a_-(t) \\ a_+(t) \\ a_{11}(t) \\ \end{array} \right). \nonumber
\end{eqnarray}
In the above expression we have introduced effective Rabi frequencies, corresponding to transitions between the states, given by expression (\ref{2}):
\begin{eqnarray}\label{4}
\begin{split}
&\Omega_1(t)=\frac{\varepsilon_1(t)(d_{10}^1-d_{10}^2)}{\sqrt{2}}, \\
&\Omega_2(t)=\frac{\varepsilon_2(t)(d_{10}^1+d_{10}^2)}{\sqrt{2}}, \\
&\Omega_3(t)=-\frac{\varepsilon_3(t)(d_{10}^1-d_{10}^2)}{\sqrt{2}}, \\
&\Omega_4(t)=\frac{\varepsilon_4(t)(d_{10}^1+d_{10}^2)}{\sqrt{2}}.
\end{split}
\end{eqnarray}
For simplicity reasons we assume below the Rabi frequencies to be real. Additionally, we have imposed the following condition on the detunings:
\begin{equation}
\Delta_1+\Delta_3=\Delta_2+\Delta_4.
\end{equation}

Note that the system of two intercoupled qubits effectively corresponds to a four level scheme, which, however possesses some additional properties. Namely, the states $\ket{\psi_{00}}$ and $\ket{\psi_{11}}$ are decoupled from one another, as the states $\ket{\psi_{+}}$ and $\ket{\psi_{-}}$ are. This is due to the fact that the corresponding effective transition dipole moments are equal to zero: $\bra{\psi_{00}}(d^1+d^2)\ket{\psi_{11}}=\bra{\psi_{-}}(d^1+d^2)\ket{\psi_{+}}=0$. Additionally, as can be seen from Eq.~(\ref{4}), for the case of identical qubits, or for ones with equal dipole moments, the state $\ket{\psi_{-}}$ is decoupled from all the others. The obtained level diagram can be considered as a Pythagorean coupling scheme with nearest-neighbor transitions, which was analyzed from a geometrical point of view in Ref.~\cite{pyth}. The scheme is also known as a double-$\Lambda$ system \cite{diamond}.

\subsection{Three intercoupled qubits}

The eigenvectors of a system of three qubits, coupled by means of exchange-type interaction, given by $\mathcal{H}_{qq}$, are the following eight states \cite{me22, me3, wang}:
\begin{eqnarray}
\begin{split}
&|\psi_{000}\rangle=|000\rangle\\
&|\psi_\mathrm{1}^\mathrm{W}\rangle=\frac{1}{\sqrt{3}}\left(|001\rangle+|010\rangle+|100\rangle\right)&\\
&|\psi_\mathrm{1}^q\rangle=\frac{1}{\sqrt{3}}\left(q|001\rangle+q^2|010\rangle+|100\rangle\right)&\\
&|\psi_\mathrm{1}^{q^2}\rangle=\frac{1}{\sqrt{3}}\left(q^2|001\rangle+q|010\rangle+|100\rangle\right)&\\
&|\psi_\mathrm{2}^\mathrm{W}\rangle=\frac{1}{\sqrt{3}}\left(|110\rangle+|101\rangle+|011\rangle\right)&\\
&|\psi_\mathrm{2}^q\rangle=\frac{1}{\sqrt{3}}\left(q|110\rangle+q^2|101\rangle+|011\rangle\right)&\\
&|\psi_\mathrm{2}^{q^2}\rangle=\frac{1}{\sqrt{3}}\left(q^2|110\rangle+q|101\rangle+|011\rangle\right)&\\
&|\psi_{111}\rangle=|111\rangle, \label{5}
\end{split}
\end{eqnarray}
with eigenenergies given as:
\begin{eqnarray}
\begin{split}
&E_{000}=-\frac{3 \omega_0}{2}; \quad
E_\mathrm{1}^W=2\lambda-\frac{\omega_0}{2};\\
&E_\mathrm{1}^q=E_\mathrm{1}^{q^2}=-\lambda-\frac{\omega_0}{2}; \quad  E_\mathrm{2}^W=2\lambda+\frac{\omega_0}{2};\\
&E_\mathrm{2}^q=E_\mathrm{2}^{q^2}=-\lambda+\frac{\omega_0}{2};
\quad E_{111}=\frac{3\omega_0}{2}.\label{6}
\end{split}
\end{eqnarray}
We have chosen the eigenvectors in the degenerate subspace such that they are simultaneously eigenstates of the cyclic shift operator with eigenvalues $q$ and $q^2$ (thus the notations $|\psi_\mathrm{1,2}^{q}\rangle$ and $|\psi_\mathrm{1,2}^{q^2}\rangle$), with $q=e^{i2\pi/3}$.

\begin{figure}[h!]
\begin{center}
\includegraphics[width=6cm]{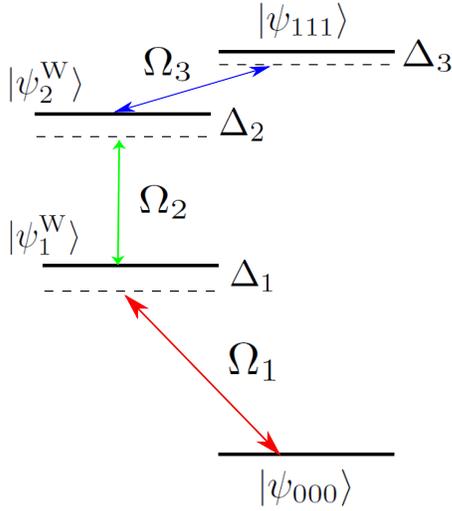}
\caption {(Color online) The effective level scheme for three intercoupled qubits, interacting with three incident laser fields, in the strong qubit-qubit coupling regime. \label{levels4}}
\end{center}
\end{figure}

As we intend to generate W and GHZ states, our aim here is to reduce the system of three intercoupled qubits to an effective four level system,
interacting with three incident laser fields and involving only the states $|\psi_\mathrm{1}^\mathrm{W}\rangle$, $|\psi_\mathrm{2}^\mathrm{W}\rangle$,
$|\psi_{000}\rangle$ and $|\psi_{111}\rangle$ (see Fig.~\ref{levels4}). For achieving this, one has to support a large enough energy gap between the states
$|\psi_\mathrm{1}^\mathrm{W}\rangle$ and $|\psi_\mathrm{1}^q\rangle$ ($|\psi_\mathrm{1}^{q^2}\rangle$) on the one hand, and between $|\psi_\mathrm{2}^\mathrm{W}\rangle$ and
$|\psi_\mathrm{2}^q\rangle$ ($|\psi_\mathrm{2}^{q^2}\rangle$) on the other hand.
As $E_\mathrm{1}^\mathrm{W}-E_\mathrm{1}^q=E_\mathrm{1}^\mathrm{W}-E_\mathrm{1}^{q^2}=E_\mathrm{2}^\mathrm{W}-E_\mathrm{2}^q=E_\mathrm{2}^\mathrm{W}-E_\mathrm{2}^{q^2}=3\lambda$,
the condition of a laser pulse to be resonant to the transition
$|\psi_{000}\rangle\leftrightarrow |\psi_\mathrm{1}^\mathrm{W}\rangle$ ($|\psi_{111}\rangle\leftrightarrow |\psi_\mathrm{2}^\mathrm{W}\rangle$),
but off resonant to the transition $|\psi_{000}\rangle\leftrightarrow |\psi_\mathrm{1}^q\rangle, |\psi_\mathrm{1}^{q^2}\rangle$
($|\psi_{111}\rangle\leftrightarrow |\psi_\mathrm{2}^q\rangle, |\psi_\mathrm{2}^{q^2}\rangle$) reads: $\lambda\sim\omega_0$. In other words,
the scheme depicted in Fig.~\ref{levels4} holds true, if one works in the strong (qubit-qubit) intercoupling regime.
In this case, the time evolution of amplitudes $\{a_{000}(t), a_{1}^\mathrm{W}(t), a_{2}^\mathrm{W}(t), a_{111}(t)\}$ is governed by means of the following set of equations:

\begin{eqnarray} \label{7}
&& i\frac{d}{d t} \left( \begin{array}{l} a_{000}(t) \\ a_{1}^\mathrm{W} \\ a_{2}^\mathrm{W} \\ a_{111}(t) \\ \end{array} \right)=\\ &&\left(
                         \begin{array}{llll}
                           0            & -\Omega_1^*(t) & 0                  & 0 \\
                           -\Omega_1(t) & \Delta_1       & -\Omega_2^*(t)     & 0 \\
                           0            & -\Omega_2(t)   & \Delta_1+\Delta_2  & -\Omega_3^*(t) \\
                           0            & 0              & -\Omega_3(t)       & \Delta_1+\Delta_2+\Delta_3 \\
                         \end{array}
                       \right) \cdot \left( \begin{array}{l} a_{000}(t) \\ a_{1}^\mathrm{W} \\ a_{2}^\mathrm{W} \\ a_{111}(t) \\ \end{array} \right), \nonumber
\end{eqnarray}
with corresponding effective Rabi frequencies:
\begin{eqnarray}\label{8}
\begin{split}
&\Omega_1(t)=\frac{\varepsilon_1(t)(d_{10}^1+d_{10}^2+d_{10}^3)}{\sqrt{3}}, \\
&\Omega_2(t)=\frac{2\varepsilon_2(t)(d_{10}^1+d_{10}^2+d_{10}^3)}{\sqrt{3}}, \\
&\Omega_3(t)=\frac{\varepsilon_3(t)(d_{10}^1+d_{10}^2+d_{10}^3)}{\sqrt{3}}, \\
\end{split}
\end{eqnarray}
where $\varepsilon_i(t)$ and $d_{kl}^i=\bra{k}d^i\ket{l}$ have the same meaning as in Eq.~(\ref{4}). This effective scheme can be also considered as a four level ladder system \cite{coherence}. We note, however, that in a general case (arbitrary strength of interqubit exchange interaction) coupling with laser fields is possible only for transitions, where the number of excited qubits changes for $\pm 1$.

\section{Generation of entangled states}\label{ent}

In the present section we propose a method for generating three classes of entangled states. Namely, we present schemes for creating Bell, W and GHZ states in described above systems of mutually coupled two and three qubits, interacting with incident laser fields. For that we use the pulse area and adiabatic passage techniques.
\subsection{Bell states}
As is known, Bell states form a basis of maximally entangled states in a two-qubit Hilbert space, and are widely used in various aspects of quantum information science. Bellow we present methods for generation of these states from factorable (non-entangled) states $\ket{00}$ (or $\ket{11}$). More precisely, starting from a state, where qubits, intercoupled by means of exchange-type interaction given by $\mathcal{H}_{qq}$ [Eq.~(\ref{1})], are in their ground state $\ket{\psi_{00}}$ (the parameters can be always chosen such that $\ket{\psi_{00}}$ is the ground state), we aim at creating the following maximally entangled states:
\begin{eqnarray}\label{9}
\begin{split}
& \ket{\varphi_{\pm}}=\frac{1}{\sqrt{2}}(\ket{00}\pm\ket{11}), \\
& \ket{\psi_{\pm}}=\frac{1}{\sqrt{2}}(\ket{10}\pm\ket{01}).
\end{split}
\end{eqnarray}

Note that two of the above states, namely, $\ket{\psi_{+}}$ and $\ket{\psi_{-}}$, are eigenstates of the initial system of mutually coupled qubits [see Eq.~(\ref{2})]. Thus,
it is possible to obtain these two states from the ground state $\ket{\psi_{00}}=\ket{00}$ by means of only one laser pulse (e.g., with a constant amplitude and of an appropriate duration $T$),
resonant to the transition $\ket{\psi_{00}}\leftrightarrow\ket{\psi_+}$ (for generating $\ket{\psi_+}$) or to the transition $\ket{\psi_{00}}\leftrightarrow\ket{\psi_-}$
(for generating $\ket{\psi_-}$). However, in order to support only one of these resonances, we have to impose a large enough energy gap between the states $\ket{\psi_{+}}$ and $\ket{\psi_{-}}$.
Since $E_+-E_-=2\lambda$, this can be achieved in the strong mutual coupling regime ($\lambda\sim\omega_0$). When this condition is satisfied, only two amplitudes are involved in the time
evolution, depending on what transition the laser pulse is resonant to:
\begin{eqnarray} \label{10}
i\frac{d}{d t} \left( \begin{array}{c} a_{00}(t) \\ a_-(t) \end{array} \right)=&& \left(
                         \begin{array}{ll}
                           0              & -\Omega_{1}(t)  \\
                           -\Omega_{1}(t) & \Delta_1              \\

                         \end{array}
                       \right) \cdot \left( \begin{array}{c} a_{00}(t) \\ a_-(t)  \\ \end{array} \right)
\end{eqnarray}
for a resonant $\ket{\psi_{00}}\leftrightarrow\ket{\psi_-}$ transition and
\begin{eqnarray} \label{11}
i\frac{d}{d t} \left( \begin{array}{c} a_{00}(t) \\ a_+(t) \end{array} \right)=&& \left(
                         \begin{array}{ll}
                           -\Delta_2     & -\Omega_{2}(t)  \\
                           -\Omega_{2}(t) & 0              \\

                         \end{array}
                       \right) \cdot \left( \begin{array}{c} a_{00}(t) \\ a_+(t)  \\ \end{array} \right)
\end{eqnarray}
for a resonant transition $\ket{\psi_{00}}\leftrightarrow\ket{\psi_+}$. Preparing initially two qubits in their ground states, we can obtain Rabi oscillation between $\ket{\psi_{00}}$ and $\ket{\psi_-}$ on the one hand [Eq.~(\ref{10})] and between $\ket{\psi_{00}}$  and $\ket{\psi_+}$ on the other hand [Eq.~(\ref{11})]. Thus, choosing the pulse area of the laser to be $\pi/2$, and working at exact resonance ($\Delta_1=\Delta_2=0$), we perform a complete population transfer from a non-entangled ground state $\ket{\psi_{00}}$ to a maximally entangled Bell state $\ket{\psi_-}$ ($\ket{\psi_+}$). We also note that the required population transfer in systems defined by Eqs.~(\ref{10}) and (\ref{11}) can be also achieved by means of the RAP (rapid adiabatic passage) technique, when the detuning $\Delta_1$ (or $\Delta_2$) is time-dependent and is changed adiabatically in such a way that $\Delta_i(\pm \infty)/\Omega_i(\pm \infty)=\pm \infty$ ($i=1, 2$). This method, unlike Rabi oscillations, is robust against the laser intensity, detuning, and interaction time variations.

On the other hand, for preparing the states $\ket{\varphi_{\pm}}$ one has to consider the full system of two qubits interacting with four laser fields, since the states $\ket{\varphi_{\pm}}$
are not eigenvectors of the Hamiltonian $\mathcal{H}_{qq}$. In other words, our target here is to generate a coherent superposition of the states $\ket{\psi_{00}}$ and $\ket{\psi_{11}}$. We additionally assume that $\Omega_1(t)=\Omega_2(t)\equiv \Omega_1(t)$ and $\Omega_3(t)=\Omega_4(t)\equiv \Omega_3(t)$. Note that in this case the Hamiltonian that governs the evolution of the amplitudes
$\{a_{00}(t), a_-(t), a_+(t), a_{11}(t)\}$ possesses one dark state among the following two:
\begin{eqnarray} \label{12}
\begin{split}
&\ket{\psi^D_1}=\sin{\theta}(t)\ket{00}-\cos{\theta}(t)\ket{11}, \\
&\ket{\psi^D_2}=\frac{1}{\sqrt{2}}(\ket{\psi_+}-\ket{\psi_-})=\ket{10}.
\end{split}
\end{eqnarray}
The first state can be realized in the two-photon resonance case $\Delta_1+\Delta_3=0$ and $\Delta_2+\Delta_4=0$ [$\tan{\theta(t)}=\Omega_3(t)/\Omega_1(t)$], while the second one is found at $\Delta_1=\Delta_2$ and $\Delta_3=\Delta_4$, i.e., when the field dressed states $\ket{\psi_{+}}$ and $\ket{\psi_{-}}$ are degenerate. As is seen from Eq.~(\ref{12}), the state $\ket{\psi^D_1}$ allows one to perform population transfer from the state $\ket{00}$ to the state $\ket{11}$ within the STIRAP technique. Meanwhile, if atoms are initially prepared in the state $\ket{10}$, the realization of $\ket{\psi^D_2}$ (for $\Delta_1=\Delta_2$ and $\Delta_3=\Delta_4$) forces the system to remain trapped there, even in the presence of adiabatically evolving fields.

\begin{figure}[ht]
\begin{center}
\small(a) \includegraphics[width=7cm]{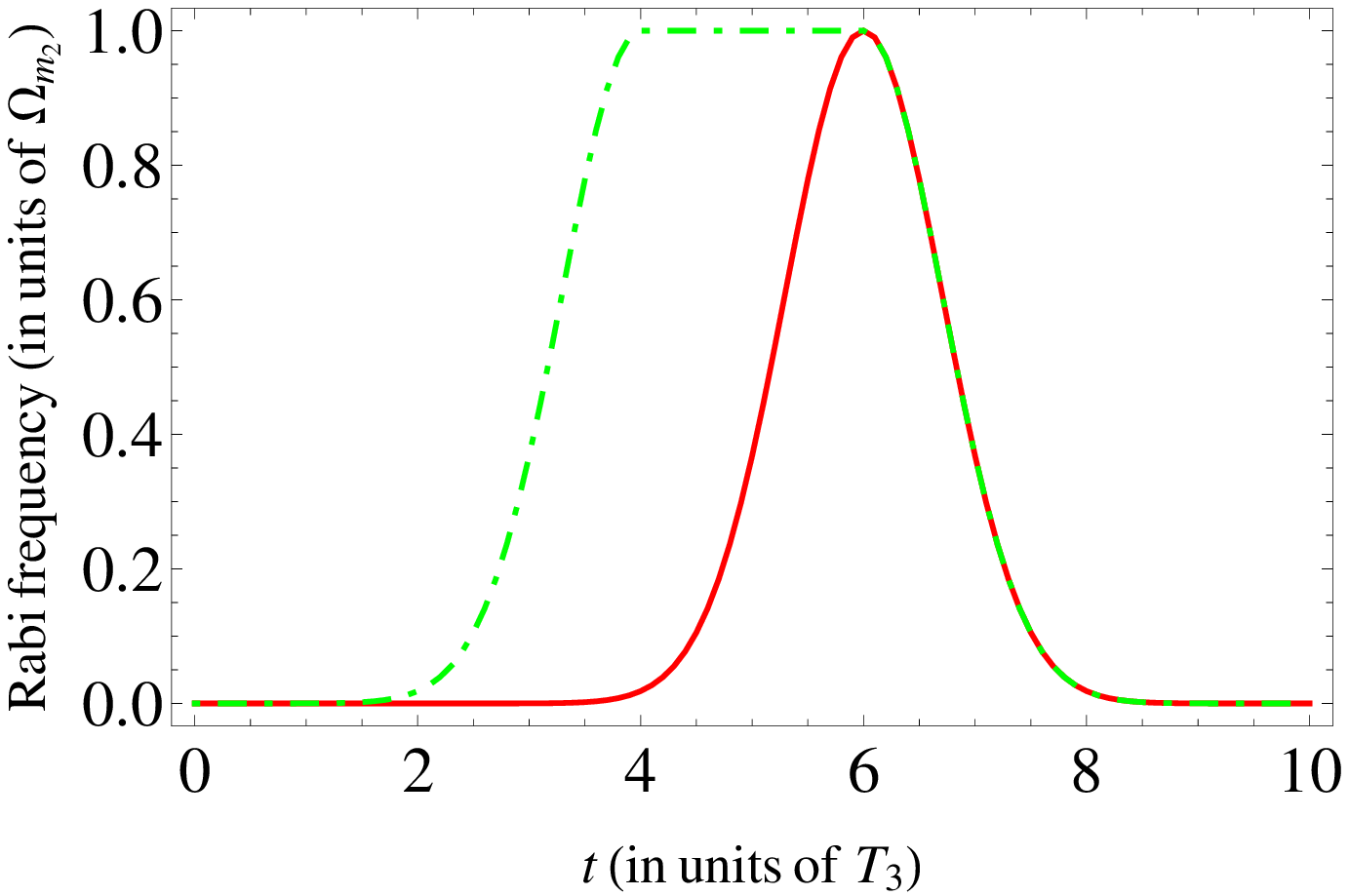}\\
\small(b) \includegraphics[width=7cm]{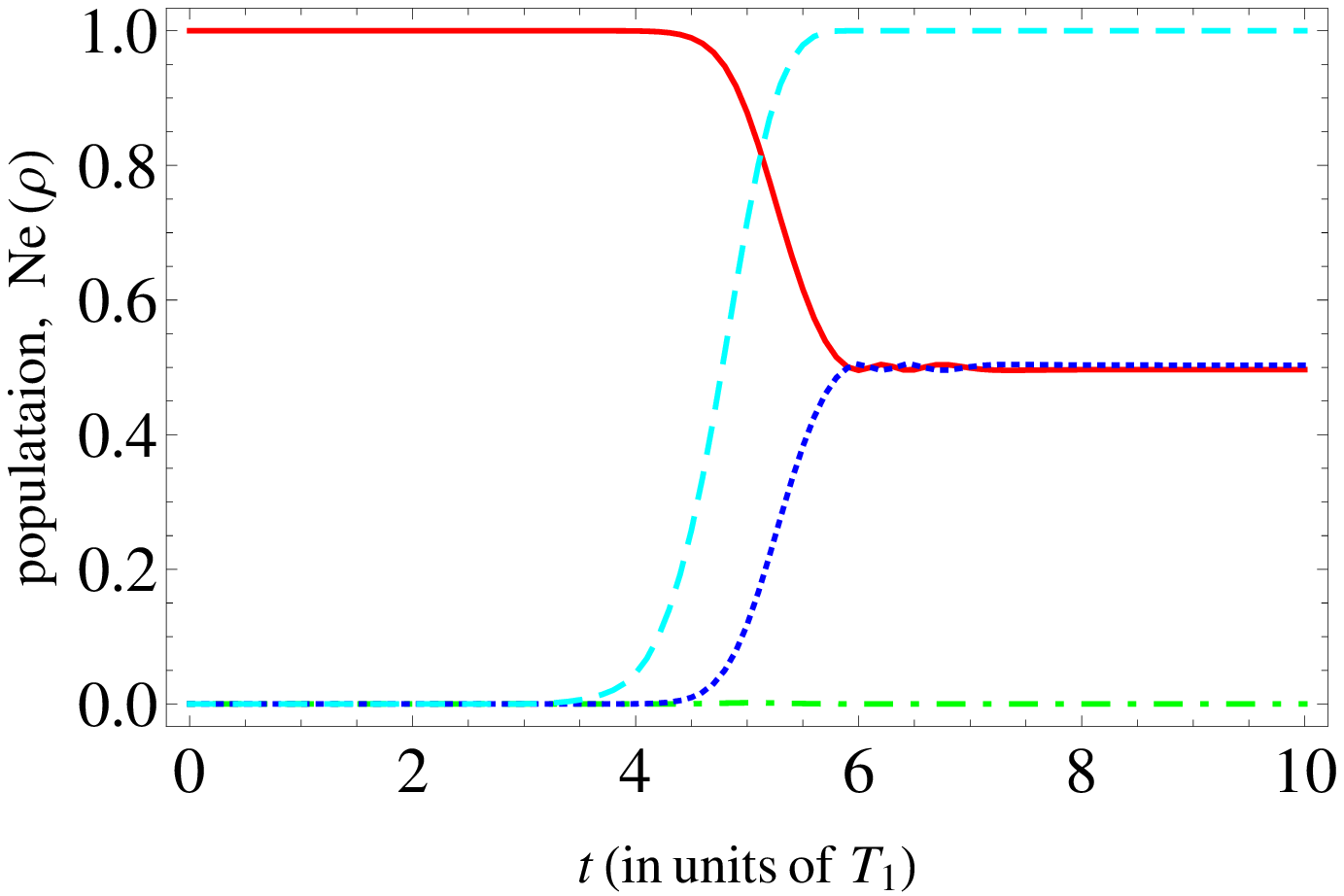}
\caption {(Color online) \small{(a) Time sequence of the pulses $\Omega_1(t)$ (full red curve) and $\Omega_3(t)$ (dotted-dashed green curve) aimed at creating the state $\ket{\varphi_-}$ in a system of two intercoupled qubits governed by the Hamiltonian $\mathcal{H}_{qq}$ [Eq.~(\ref{1})]. Here $\Omega_{m_1}T_3=\Omega_{m_3}T_3=7.5$, $\tau_1/T_3=6$, $\tau_3/T_3=4$ and $\Delta_1=\Delta_2=\Delta_3=\Delta_4=0$;
(b) Time evolution of eigenstate populations (full red curve: $\ket{\psi_{00}}$, dotted-dashed green curve: $\ket{\psi_{+}}$ and $\ket{\psi_{-}}$,
dotted blue curve: $\ket{\psi_{11}}$) and the logarithmic negativity $Ne(\rho)$ (dashed cyan curve). \label{rstirap}}}
\end{center}
\end{figure}

Furthermore, one can use the fractional STIRAP method, for generating $\ket{\varphi_-}$ from the initial ground state $\ket{00}$. The mixing angle $\theta(t)$ should evolve here in such a way that $\theta(-\infty)=0$ and $\theta(+\infty)=\pi/4$. This is achieved by the application of a laser pulse $\Omega_1(t)=\Omega_{m_1}e^{-(t-\tau_1)^2/T_1^2}$, overlapping with $\Omega_3(t)$, defined as (see Fig.~\ref{rstirap}):
\begin{equation}
\Omega_3(t)=\left
\{\begin{array}{l}
\Omega_{m_3} e^{-(t-\tau_3)^2/T_3^2}, \; \mathrm{if} \; t<\tau_3 \\
\Omega_{m_3}, \; \mathrm{if} \; \tau_3<t<\tau_1                  \\
\Omega_{m_3} e^{-(t-\tau_1)^2/T_3^2}, \; \mathrm{if} \; t>\tau_1.
\end{array} \right.\label{om3}
\end{equation}
These two Gaussian laser pulses perform a partial atomic population transfer from the state $\ket{00}$ to the state $\ket{11}$, eventually resulting in $\ket{\varphi_-}$. For quantifying the amount of entanglement during time evolution of the system, we have used the logarithmic negativity $Ne(\rho)$, which is defined as
\begin{equation}
Ne(\rho)=\log_2\left|\left|\rho^{T_A}\right|\right|, \label{14}
\end{equation}
where $\left|\left|\rho^{T_A}\right|\right|$ is the trace norm of the partial transposed $\rho^{T_A}$ of a bipartite density matrix $\rho=\ket{\Psi}\bra{\Psi}$.

Similarly, the state $\ket{\varphi_+}$ can be also constructed by the fractional STIRAP method: imposing conditions for the realization of the dark state $\ket{\psi^D_1}$ (i.e., working in the regime of a two-photon resonance), and starting from the ground state $\ket{00}$, one can obtain the entangled state $\ket{\varphi_+}$, if the fields $\Omega_1(t)$ and $\Omega_3(t)$ have a relative $\pi$ phase shift.

Another method for generating the state $\ket{\varphi_-}$ (as well as $\ket{\varphi_+}$) is the pulse area method. Assuming that the system is initially in its non-entangled ground state, i.e., $\{a_{00}(0)=1, a_-(0)=0, a_+(0)=0, a_{11}(0)=0\}$ and that $\Omega_1(t)=\Omega_2(t)\equiv\Omega_1(0)$, $\Omega_3(t)=\Omega_4(t)\equiv\Omega_3(0)$, the solution of Eq.~(\ref{3}) takes the following form ($\Delta_1=\Delta_2=\Delta_3=\Delta_4=0$):
\begin{eqnarray} \label{13}
\begin{split}
&a_{00}=\cos \left( \sqrt{2(\Omega _{1_0}^2+\Omega _{3_0}^2)}t\right)\frac{\Omega _{1_0}^2 }{\Omega _{1_0}^2+\Omega
   _{3_0}^2}+\frac{\Omega _{3_0}^2}{\Omega _{1_0}^2+\Omega _{3_0}^2}, \\
&a_{+,-}=i \sin \left(\sqrt{2(\Omega _{1_0}^2+\Omega _{3_0}^2)}t\right)\frac{ \Omega _{1_0} }{\sqrt{2(\Omega _{1_0}^2+\Omega _{3_0}^2)}}, \\
&a_{11}=\cos \left(\sqrt{2(\Omega_{1_0}^2+\Omega_{3_0}^2)}t\right)\frac{\Omega _{1_0} \Omega _{3_0 }}{\Omega _{1_0}^2+\Omega _{3_0}^2}-\frac{\Omega _{1_0} \Omega _{3_0}}{\Omega _{1_0}^2+\Omega _{3_0}^2}.
\end{split}
\end{eqnarray}

\begin{figure}[h!]
\begin{center}
\includegraphics[width=7cm]{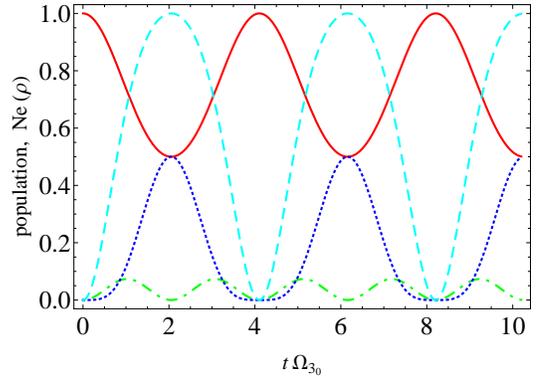}
\caption {(Color online) Time evolution of the eigenstate populations (full red curve: $\ket{\psi_{00}}$, dotted-dashed green curve: $\ket{\psi_{+}}$ and $\ket{\psi_{-}}$, dotted blue curve: $\ket{\psi_{11}}$) and the logarithmic negativity $Ne(\rho)$ (dashed cyan curve) of a system of two intercoupled qubits governed by the Hamiltonian $\mathcal{H}_{qq}$ [Eq.~(\ref{1})], interacting with four laser pulses having constant amplitudes $\Omega_{i_0}=\Omega_{i}(0)$ ($i=1,2,3,4$). Here $\Omega_{1_0}/\Omega_{3_0}=\Omega_{2_0}/\Omega_{3_0}=1-\sqrt{2}$, $\Omega_{4_0}/\Omega_{3_0}=1$ and $\Delta_1=\Delta_2=\Delta_3=\Delta_4=0$. \label{phi}}
\end{center}
\end{figure}

The above set of equations shows that pulses of a duration $\sqrt{2(\Omega_{1_0}^2+\Omega_{3_0}^2)} T=\pi$ generate the state $\ket{\varphi_\pm}$, if the Rabi frequencies satisfy the condition $\Omega_{1_0} T= \pm (1-\sqrt{2}\Omega_{3_0}) T$. Corresponding Rabi oscillations are shown in Fig.~\ref{phi}.

\begin{figure}[h!]
\begin{center}
\includegraphics[width=7cm]{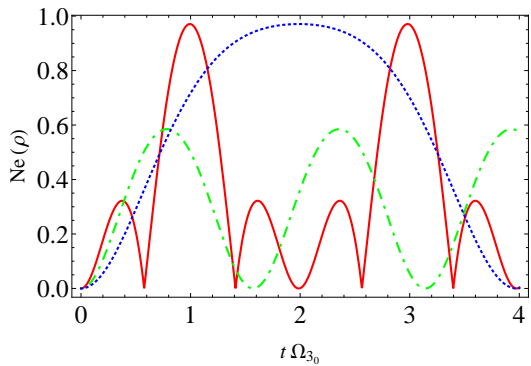}
\caption {(Color online) Time evolution of the logarithmic negativity $Ne(\rho)$ of a system of two intercoupled qubits governed by the Hamiltonian $\mathcal{H}_{qq}$ [Eq.~(\ref{1})], interacting with four laser pulses having constant amplitudes $\Omega_{i_0}=\Omega_{i}(0)$ ($i=1,2,3,4$). Here $\Omega_{4_0}/\Omega_{3_0}=1$ and $\Omega_{1_0}/\Omega_{3_0}=\Omega_{2_0}/\Omega_{3_0}=2$ (full red curve), $\Omega_{1_0}/\Omega_{3_0}=\Omega_{2_0}/\Omega_{3_0}=1$ (dotted-dashed green curve), $\Omega_{1_0}/\Omega_{3_0}=\Omega_{2_0}/\Omega_{3_0}=1/2$ (dotted blue curve); $\Delta_1=\Delta_2=\Delta_3=\Delta_4=0$. \label{neg_2}}
\end{center}
\end{figure}

We note that the proposed technique allows one to generate not only Bell states, when starting from a separable ground state,
but also states with a different amount of entanglement. Figure~\ref{neg_2} shows the time evolution of the logarithmic negativity, for different relations between $\Omega_{1_0}$ and $\Omega_{3_0}$.

Furthermore, one can manipulate the amount of entanglement of the system in a continuous way, by choosing a corresponding pulse area.

Although the above discussed adiabatic and pulse area methods for generating the states $\ket{\varphi_-}$ and $\ket{\varphi_+}$ do not involve explicitly the condition of a strong interqubit coupling, it is worth to note that a small energy gap between the states $\ket{\psi_{+}}$ and $\ket{\psi_{-}}$ may result in bichromatic effects, where the standard RWA cannot be applied \cite{bichrom}. In particular, distinct laser fields could not be assigned to a unique transition, what brings about the so-called {\it ambiguous coupling} \cite{ambig}. Thus  for the implementation of the above schemes one still requires a relatively strong qubit-qubit coupling. However, the ambiguity can be removed if only one laser is coupled to the system. Although this would result in non-vanishing detunings (if the pulse is resonant to the transition $\ket{\psi_{00}}\leftrightarrow \ket{\psi_{-}}$, it is not resonant to $\ket{\psi_{00}}\leftrightarrow \ket{\psi_{+}}$ anymore), the adiabatic methods still remain operational due to their robustness.

\subsection{W and GHZ states}

Equation~\ref{5} shows that a system of three coupled qubits interacting by means of exchange-type interaction possesses the W state ($\ket{\psi_1^W}$) as an eigenstate. Thus, starting from a separable ground state $\ket{000}$ and making use of the pulse area technique with only one laser pulse, resonant to the transition $\ket{\psi_{000}}\leftrightarrow\ket{\psi_{1}^W}$, we can perform a complete population transfer to the state $\ket{\psi_{1}^W}$ ($\Delta_1=0$). As in the previous subsection we work here in the strong qubit-qubit coupling regime, i.e., $\lambda \sim \omega_0$. In this case only the populations of the states $\ket{\psi_{000}}$ and $\ket{\psi_{1}^W}$ change in time, while the other states remain empty. Thus, a $\pi/2$ pulse performs a complete population transfer from $\ket{\psi_{000}}$ to $\ket{\psi_{1}^W}$, as discussed previously. Additionally, this population transfer can be also performed by means of a more robust method, the rapid adiabatic passage (RAP), which involves a time-dependent detuning $\Delta_1(t)$. It is also worth to note that the state $\ket{\psi_{2}^W}$, being another W state, present in our model, can be readily obtained from the separable ground state $\ket{\psi_{000}}$ by means of the conventional STIRAP method, within a counterintuitive sequence of the pulses $\Omega_1(t)$ and $\Omega_2(t)$.

\begin{figure}[t]
\begin{center}
\small(a) \includegraphics[width=7cm]{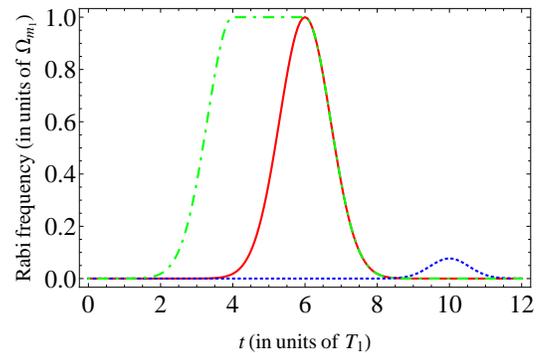}\\
\small(b) \includegraphics[width=7cm]{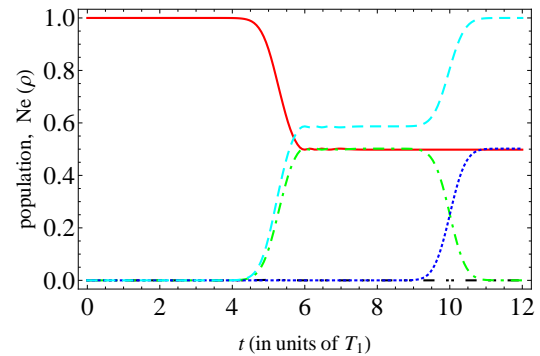}\\
\caption {(Color online) \small{(a) Time sequence of the pulses $\Omega_1(t)$ (full red curve), $\Omega_2(t)$ (dotted-dashed green curve) and $\Omega_3(t)$ (dotted blue curve) aimed at creating a three-qubit GHZ state in a system of three intercoupled qubits in the strong mutual coupling regime. Here $\Omega_{m_1}T_1=\Omega_{m_2}T_1=15$ $\Omega_{m_3} T_1=1.1535$, $T_2/T_1=1$, $T_3/T_1\approx 0.77$, $\tau_1/T_1=6$, $\tau_2/T_1=4$, $\tau_3/T_1=10$ and $\Delta_1=\Delta_2=\Delta_3=0$; (b) Time evolution of eigenstate populations (full red curve: $\ket{\psi_{000}}$,  double dotted-dashed black curve: $\ket{\psi_{1}^W}$, dotted-dashed green curve: $\ket{\psi_{2}^W}$, dotted blue curve: $\ket{\psi_{111}}$) and the logarithmic negativity $Ne(\rho)$  (dashed cyan curve). \label{pulses}}}
\end{center}
\end{figure}

For generating a three-qubit GHZ state we use a combination of the fractional STIRAP and the $\pi$ pulse techniques. More precisely, we choose pulses of Gaussian shape: $\Omega_1(t)=\Omega_{m_1}e^{-(t-\tau_1)^2/T_1^2}$, $\Omega_3(t)=\Omega_{m_3}e^{-(t-\tau_3)^2/T_3^2}$ and
\begin{equation}
\Omega_2(t)=\left
\{\begin{array}{l}
\Omega_{m_2} e^{-(t-\tau_2)^2/T_3^2}, \; \mathrm{if} \; t<\tau_2 \\
\Omega_{m_2}, \; \mathrm{if} \; \tau_3<t<\tau_1                  \\
\Omega_{m_2} e^{-(t-\tau_1)^2/T_3^2}, \; \mathrm{if} \; t>\tau_1,
\end{array} \right.
\end{equation}
and apply them in the sequence depicted in Fig.~\ref{pulses}(a). This sequence drives a part of the atomic population out of $\ket{\psi_{000}}$ to the state $\ket{\psi_{2}^W}$, which is afterwards directed to $\ket{\psi_{111}}$ by means of a $\pi$ pulse. As a result, we generate the state $a\ket{000}+b\ket{111}$, with $a\approx b\approx 1/\sqrt{2}$. For having $a\approx b$, one has also to provide a large enough time delay $\tau_3$ of the $\pi$ pulse, that additionally assures maximal coherence of atomic populations \cite{coherence}. Note that for quantifying the amount of entanglement during this process by means of the logarithmic negativity, one has to perform a partial trace out operation over one of the qubits. However, the index number of the traced out qubit can be chosen arbitrarily here, as the system possesses a translational symmetry.

\begin{figure}[t]
\begin{center}
\small(a) \includegraphics[width=7cm]{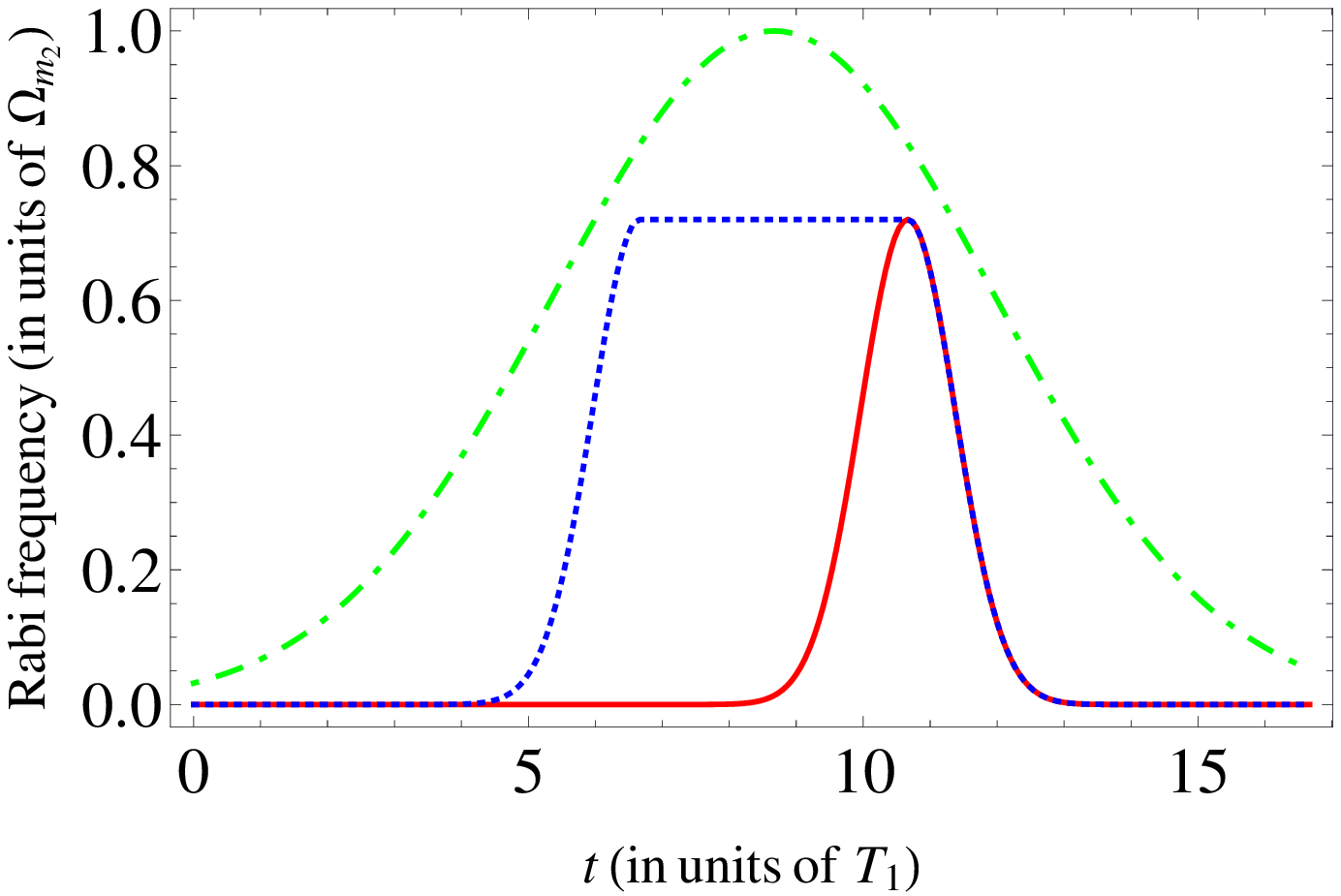}\\
\small(b) \includegraphics[width=7cm]{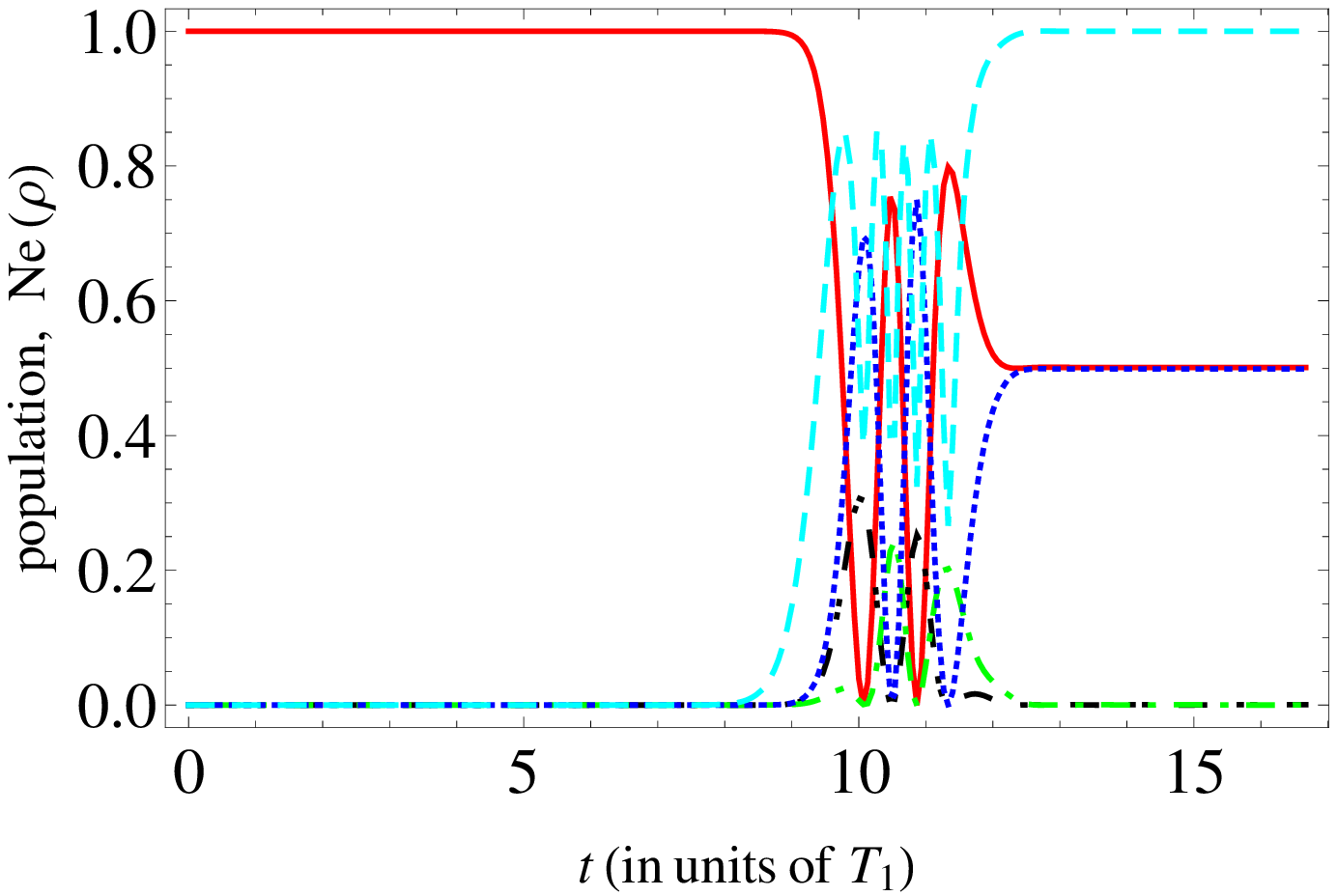}\\
\caption {(Color online) \small{(a) Time sequence of the pulses $\Omega_1(t)$ (full red curve), $\Omega_2(t)$ (dotted-dashed green curve) and $\Omega_3(t)$ (dotted blue curve) aimed at creating a three-qubit GHZ state in a system of three intercoupled qubits in the strong mutual coupling regime. Here $\Omega_{m_1}T_1=\Omega_{m_3}T_1=7.5$ $\Omega_{m_2} T_2=48.615$, $T_2/T_1\approx4.67$, $T_3/T_1=1$, $\tau_1/T_1\approx 10.67$, $\tau_2/T_1\approx 8.67$, $\tau_3/T_1\approx 6.67$ and $\Delta_1=\Delta_2=\Delta_3=0$; (b) Time evolution of eigenstate populations (full red curve: $\ket{\psi_{000}}$, double dotted-dashed black curve: $\ket{\psi_{1}^W}$, dotted-dashed green curve: $\ket{\psi_{2}^W}$, dotted blue curve: $\ket{\psi_{111}}$) and the logarithmic negativity $Ne(\rho)$  (dashed cyan curve). \label{new}}}
\end{center}
\end{figure}

Finally, the desired superposition of $\ket{\psi_{000}}$ and $\ket{\psi_{111}}$ can be also obtained without the application of a $\pi$ pulse. The idea is to perform a fractional STIRAP from $\ket{\psi_{000}}$ to $\ket{\psi_{111}}$ by means of $\Omega_1(t)=\Omega_{m_1}e^{-(t-\tau_1)^2/T_1^2}$ and $\Omega_3(t)$, which has the same definition as in Eq.~(\ref{om3}), with a laser pulse $\Omega_2(t)=\Omega_{m_2}e^{-(t-\tau_2)^2/T_2^2}$ switched on for the intermediate transition $\ket{\psi_{1}^W}\leftrightarrow\ket{\psi_{2}^W}$ during the whole interaction time. The efficiency of the technique is shown in Fig.~\ref{new}.

\section{Conclusion and Discussion}\label{conc}

We have presented several schemes for generating three types of entangled (Bell, GHZ and W) states by means of adiabatic and pulse area methods in systems of intercoupled qubits. We show that  in the strong qubit-qubit coupling regime, within the Rabi oscillation or rapid adiabatic passage techniques, two of the Bell states, namely, the triplet and the singlet states (being the eigenstates of the bare system of two intercoupled qubits), can be obtained from a separable ground state by making use of only one laser pulse. On the other hand, generation of the remaining two Bell states in the full system requires the strong intercoupling strength for avoiding bichromatic effects, and can be performed within the pulse area and fractional STIRAP methods. Additionally, one can manipulate the amount of the entanglement of a system in a continuous way here, by choosing an appropriate area of laser pulses.

Furthermore, we point out that a system of three strongly coupled qubits is effectively equivalent to a four level ladder system, possessing two W and two separable states. This allows one to obtain W states from a non-entangled ground state, as well as to create a GHZ state by means of a combination of the fractional STIRAP and $\pi$ pulse techniques. It is worth to note that a system of $N$ qubits, interacting through the aforementioned exchange interaction, possesses two $N$-qubit W states, analogous to the above-described $\ket{\psi_{1}^W}$ and $\ket{\psi_{2}^W}$. Corresponding eigenenergies are given as $E_{1}^W=(N-1)\lambda-(n-2)\omega_0/2$ and $E_{2}^W=(N-1)\lambda+(n-2)\omega_0/2$. On the other hand, $N-1$ times degenerate $N$-qubit generalizations of the states $|\psi_\mathrm{1,2}^{q}\rangle$ and $|\psi_\mathrm{1,2}^{q^2}\rangle$, which we denote as $\ket{\psi_\mathrm{1,2}^{q^k}}$ ($q=e^{i2\pi/N}$ and $k=1, ..., N-1$) are also present in the system. The energy gap between these and corresponding W states is $N\cdot \lambda$, what makes the above-imposed strong qubit-qubit coupling condition less strict. However, existence of additional eigenstates may result in undesirable resonances, making the bare system more complicated. We will address this question, as well as the possibility of generalization of the presented schemes for an arbitrary number of qubits, in our future works.

Meanwhile, the interqubit exchange-type coupling can be implemented in a number of systems, as, e.g., coupled semiconductor quantum dots, superconducting phase qubits, atoms interacting with a cavity in the dispersive limit. The latter realization does not allow one to have a strong qubit-qubit coupling regime. Nevertheless, since the condition of a strong interaction becomes less strict with the increase of the number of qubits, the system can be still  used for creating entangled states, with a further macroscopic separation of two parties. This can be achieved by making non-excited atoms travel through a cavity with transverse laser beams, and by controlling the interaction time (and therefore the atomic velocity). The procedure results in a macroscopically separated output of aforementioned entangled states (see, e.g., Refs.~\cite{DLCZ, cavent}).

Finally, the effects of quantum decoherence, appearing, e.g., due to the environmental coupling with a large number of uncontrolled degrees of freedom, should be also considered here, for a thorough understanding of how the above techniques behave under realistic experimental conditions: an important issue that we will address in our future works. Nevertheless, we note that several experiments proved the possibility of implementing Rabi oscillations, relevant to our scheme, in a biexciton, confined in single GaAs and InGaAs/GaAs quantum dots \cite{expqd1, inter1}  (with dephasing times up to nanoseconds \cite{expqd}). Although these systems are not scalable beyond two qubits, they demonstrate the potential for coherent optical control in scalable architectures based on multidot systems. Another platform for the implementation of more-than-two-qubit entangling protocols appears to be systems of coupled superconducting qubits. The origin of decoherence here is, for instance, the fluctuation of external control parameters (such as gate voltages and magnetic fluxes), that can be minimized when operating upon the so-called optimal point, where the first-order noise cancels \cite{optimal}. Specifically, three-qubit entangled states of coupled Josepshon-junction qubits have been implemented recently by making use of quantum C-NOT and iSWAP gates \cite{jos2}. The relaxation and spin-echo dephasing times were shown to be of a few hundred nanoseconds here, that, along with substantially shorter gate operation times, allowed the construction of the target states with a rather high fidelity.

\section*{acknowledgements}

The authors are grateful to Eilon Poem and Joshua Nunn for useful discussions. This research was conducted within the scope of the International Associated Laboratory (CNRS-France \& SCS-Armenia) IRMAS. We acknowledge additional support from GA-ITN-214962-FASTQUAST, ERA-WIDE GA-INCO-295025-IPERA and PIRSES-GA-2013-612707-DIONICOS FP7 programs and SCS MES RA research project No. SCS 13-1C137. L.C. gratefully acknowledges the support from the Conseil R\'{e}gional de Bourgogne.

\end{document}